\documentclass[sigconf]{acmart}
\acmConference[EASE 2024]{The 28th International Conference on Evaluation and Assessment in Software Engineering}{18–21 June, 2024}{Salerno, Italy}

\AtBeginDocument{%
  \providecommand\BibTeX{{%
    \normalfont B\kern-0.5em{\scshape i\kern-0.25em b}\kern-0.8em\TeX}}}

\setcopyright{acmlicensed}
\copyrightyear{2024}
\acmYear{2024}
\acmDOI{XXXXXXX.XXXXXXX}

\begin{document}

\title{Improving classifier-based effort-aware software defect prediction by reducing ranking errors}

\author{Yuchen Guo}
\affiliation{%
    \institution{Xi'an Jiaotong University}
    \city{Xi'an}
    \country{China}}
\email{wispcat@stu.xjtu.edu.cn}

\author{Martin Shepperd}
\affiliation{%
    \institution{Brunel University London}
    \city{London}
    \country{UK}
}
\email{martin.shepperd@brunel.ac.uk}
\author{Ning Li}
\affiliation{%
    \institution{Northwestern Polytechnical University}
    \city{Xi'an}
    \country{China}
}
\email{lining@nwpu.edu.cn}

\renewcommand{\shortauthors}{Guo et al.}

\begin{abstract}
    \textbf{Context}: Software defect prediction utilizes historical data to direct software quality assurance resources to potentially problematic components. Effort-aware (EA) defect prediction prioritizes more bug-like components by taking cost-effectiveness into account. In other words, it is a ranking problem, however, existing ranking strategies based on classification, give limited consideration to ranking errors.
    \newline
    \textbf{Objective}: Improve the performance of classifier-based EA ranking methods by focusing on ranking errors.
    \newline
    \textbf{Method}: We propose a ranking score calculation strategy called EA-Z which sets a lower bound to avoid near-zero ranking errors.  We investigate four primary EA ranking strategies with 16 classification learners, and conduct the experiments for EA-Z and the other four existing strategies.\newline
    \textbf{Results}: Experimental results from 72 data sets show EA-Z is the best ranking score calculation strategy in terms of Recall@20\% and Popt when considering all 16 learners.
    For particular learners, imbalanced ensemble learner UBag-svm and UBst-rf achieve top performance with EA-Z.
    \newline
    \textbf{Conclusion}: Our study indicates the effectiveness of reducing ranking errors for classifier-based effort-aware defect prediction. We recommend using EA-Z with imbalanced ensemble learning.
    
\end{abstract}


\received{18 Jan 2024}
\received[Accepted]{6 Mar 2024}

\maketitle

\section{Introduction}

The effort involved in software quality assurance is a major part of the cost of software engineering.  Estimates of locating and repairing defects range from 40-50\% to 60-80\% of total development costs \cite{hamill2017analyzing}.  Hence it is no surprise that predicting defect-prone software components has been a major research topic for some time \cite{Hall12,malhotra2015systematic} with the goal of enabling testing to be a more focused activity.

Software defect prediction aims to help developers locate bugs and allocate testing resources more efficiently.  A popular approach to software defect prediction is machine learning which utilizes historical data to predict whether components, e.g., files, classes, methods or commits are defect-prone or not.  For such purposes, predicted outcomes are normally seen as dichotomous (either defect-prone or not) so the predictor is referred to as a classifier.  Indeed, some researchers have suggested they are capable of ``an appealing degree'' of classification accuracy \cite{lessmann2008benchmarking}.

However, machine learning-based defect prediction still faces some practical limitations. As pointed out by Arisholm et al.~\cite{arisholm2007data,arisholm2010systematic}, it assumes equal costs for detecting and repairing any defect.  But this simplification is not true in practice.  Empirical studies, such as the Hammill and Goseva-Popstojanova study of software defect repairs at NASA found that ``83\% of the total fix implementation effort was associated with only 20\% of failures'' \cite{hamill2017analyzing}.  Similarly, Kamei et al.~\cite{kamei2010revisiting} found only 20\% of test effort was needed to detect up to 74\% of all faults.  Naturally, software engineers will wish to maximize the number of defects testing per unit of effort, thus ignoring the disparity in effort to test is likely to lead to strongly sub-optimal solutions.

Consequently, a more practical approach called effort-aware (EA) software defect prediction considers software defect prediction as a sorting task~\cite{mende2010effort,kamei2013large,li2020effort,yu2024improving} instead of classification.
The optimal ordering to achieve the maximum resource efficiency is descending order of the defect/LOC ratio, or defect density~\cite{huang2019revisiting,li2023revisiting}.  By following the rank or sorting order predicted by EA models, developers can find more defects with less effort and thus improve resource efficiency.  Subsequently a number of EA methods have been proposed \cite{mende2010effort, kamei2013large, yang2017tlel,chen2018multi,huang2019revisiting,li2020effort,yu2023multi,yu2024improving}. 

Intuitively, the defect density can be calculated as ranking score from classification predictions by the defect/LOC ratio~\cite{mende2010effort,yang2017tlel,huang2019revisiting,ni2020revisiting,yu2023finding}.  However, existing classifier-based methods give limited consideration to ranking errors.  Guo et al.~\cite{Guo2018} found that there is a gap between classification performance and EA ranking performance given skewed effort distributions.  Yu et al.~\cite{yu2024improving} further pointed out that optimizing classification accuracy is not directly related to the ranking performance.
So they build a regression model EALTR, directly maximizing the proportion of the found bugs~(ProB20).  However, even with EALTR optimizing to ProB20 it still only wins over the simplest unsupervised method ManualUp\cite{zhou2018far} for 14 out of 30 datasets in terms of ProB20~\cite{yu2024improving}. 

In this paper, we choose to improve classifier-based methods since we believe explicit modeling with the defect/LOC ratio is also a part of reducing ranking errors.  Further, we identify a novel type of error-ranking for when a classifier makes non-defective predictions.  We label this problem "Minor Chaos" (and discuss it further in Section~\ref{sec:minor_chaos}).  To counteract "Minor Chaos", we propose an EA ranking score calculation strategy EA-Z, which sets a lower boundary $\zeta$ to prevent near-zero prediction in ranking scores which we argue can reduce the ranking errors due to this problem.

For the empirical part of this work, we investigate four primary ranking strategies plus our new strategy EA-Z. The ranking strategies are applied to 16 classification learners on 72 real-world data sets.  This gives a comprehensive view about the strategy how to combine EA rank prediction with trained classifiers.  The 16 classification learners including five types of common learning algorithms and two ensemble imbalanced learning methods, are detailed in Section~\ref{sec:algorithm}.  The 72 real-world data sets are drawn from four sources of software defect prediction dataset, half of them are commit-level (just-in-time) datasets and the remainder are file-level and class-level (traditional) datasets.

This paper seeks to make the following contributions.
\begin{enumerate}
    \item Recasting EA-defect prediction as a ranking rather than a classification problem enables new improvement opportunities.
    \item We propose a new method, EA-Z, to rank components to better address ranking errors.
    \item We then compare our approach with four current methods via and extensive empirical study.
    \item Our experimental results support the effectiveness of reducing ranking errors by our method EA-Z.
    \item We quantify the trade-off between Recall20 and IFA as a reference for selecting methods.
\end{enumerate}

\section{Related Work} \label{sec:RelWork}

\begin{figure}[ht]
	\includegraphics[width=0.7\columnwidth]{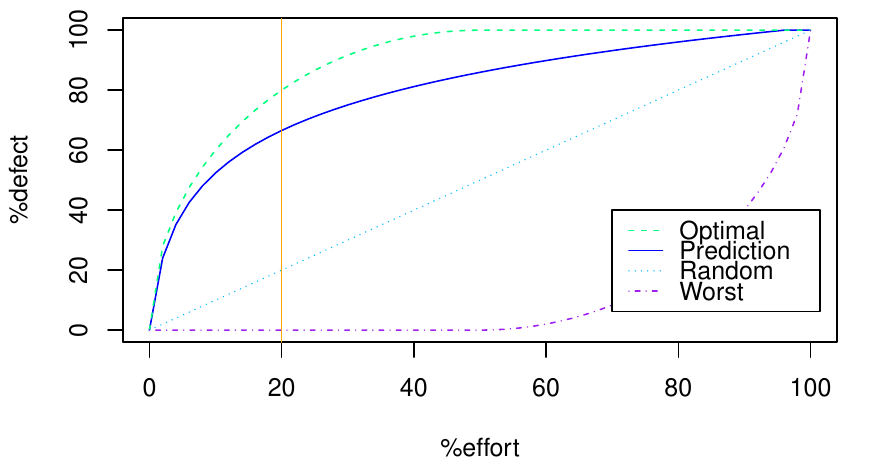}
	\caption{Cost-effectiveness Curve}
	\label{fig:curve}	
	
\end{figure}

Our starting point is the cost-effectiveness curve (see Fig.~\ref{fig:curve}) proposed by Arisholm et al.~\cite{arisholm2007data,arisholm2010systematic} as an evaluation criterion for defect prediction systems. All cost-effectiveness measures for effort-aware prediction are calculated from this curve.

The key insight is that the relationship between the proportion of defects detected and the proportion of effort expended need not be linear. 
A number of researchers have reported something approximating to a 20:80 rule where only 20\% of the effort is typically needed to address 80\% of the defects \cite{hamill2017analyzing}.

This insight makes EA software defect prediction a sorting task rather than classification.  From the perspective of making a ranking prediction, the existing effort-aware approaches can be summarized into three categories:
    \begin{enumerate}
        \item classifier-based methods;
        \item learn-to-rank methods;
        \item simple unsupervised methods.
    \end{enumerate}

The first category is classifier-based methods. As summarized by Li et al.~\cite{li2023revisiting}, there are four primary ranking score calculation strategies for classifier-based approaches: 
\begin{enumerate}
    \item Prob
    \item Label/LOC
    \item CBS+
    \item Prob/LOC
\end{enumerate}
\noindent
where Prob is predicted probability to be defective, Label is predicted label whether defective or not.

Mende and Koschke~\cite{mende2010effort} proposed the first and second listed effort-aware defect prediction models called R-ad and R-dd, where R-dd is Label/LOC.  R-ad considers defective probability as risk, then adjusts it by weight to sort defect-candidates.  R-dd calculates ranking score by label/LOC, where label is the binary prediction defective or not; LOC is the reviewing effort.  Their results show that the prediction performance of EA models improved, compared to ranking by only Prob.

Subsequently, Huang et al.~\cite{huang2017cbs} proposed two versions of the classify-before-sorting (CBS) method to improve Label/LOC, as they argue classification has higher priority than sorting.  CBS sorts faulty classes first by LOC in an inspection queue and then appends fault-free classes afterwards in increasing order of LOC.  As an improved version of CBS, CBS+~\cite{huang2019revisiting} undertakes parallel sorting by Prob/LOC, but uses the defective probability instead of predicted binary label.

Differing from Huang et al., is Yang et al.'s~\cite{yang2014slice} work using a slice-based cohesion metric, where they modified R-dd to an unnamed approach -- Prob/LOC.  Their results show the practical value of their cohesion input metric.

In addition, there are other studies that have applied Prob/LOC model by combining Prob/LOC with a specific classifier, including Prob/LOC with deep learning (Deeper)~\cite{yang2015deep},
Prob/LOC with ensemble learning (TLEL)~\cite{yang2017tlel},
Prob/LOC with semi-supervised ensemble learning (EATT)~\cite{li2020effort}.
Note that although there are five papers that have applied Prob/LOC in different ways, we view them all as a ranking score calculation strategy, since all Prob/LOC approaches suffer from the Minor Chaos problem, as described in Section~\ref{sec:minor_chaos}.

Beside the four primary ranking score calculation strategies for classifier-based methods, Qu et al.~\cite{qu2021leveraging} leveraged developer information as weights added to Label/LOC, Prob/LOC and CBS+, i.e., they improve Prob/LOC to
$\frac{Prob*NumDev^\alpha}{LOC} = (Prob/LOC)*NumDev^\alpha$.
$NumDev^\alpha$ is the developer information.  Similarly, Du et al.~\cite{du2022corebug} leveraged coreness,\footnote{Here coreness is the coreness of class c computed by k-core decomposition.} of their class dependency network to improve Prob/LOC to
$ \frac{Prob*coreness}{LOC} = (Prob/LOC)*coreness $.
Khatri and Singh~\cite{khatri2023effective} calculated ranking score by $(Prob/LOC)* avg\_cc$, where avg\_cc is the average cyclomatic complexity of the graph.  However, the rank errors caused by Minor Chaos still exist since Prob is independent in these formulae.  Moreover, adding weights to primary ranking score calculation is potentially a risk against prior knowledge of the "defect/LOC" ratio. The weights are required to be accurate and highly correlated to the optimal rank, otherwise they may lead to ranking errors.

The second category is learn-to-rank methods.  The ranking score is not calculated from classification, but by learning from the optimal rank, i.e., descending order of defect density (the defect/LOC" ratio).  Kamei et al.~\cite{kamei2013large} applied regression on defect-density to predict this order, which is named EALR (effort-aware linear regression). EALR was reported to find 35\% of defects with only 20\% effort.

More sophisticated than simple regression, some studies applied genetic algorithms or evolution algorithms to learn the effort-aware rank.  Chen et al.~\cite{chen2018multi} formalized EA prediction as a multi-objective optimization problem with two objectives: recall of defects as a benefit and effort as a cost and then applied the multi-objective learning algorithm NSGA-II  to solve it.  Likewise, Yu et al.~\cite{yu2023multi} also used NSGA-II, however as a classifier, and calculated ranking score as the same as CBS+.

Note that we do not see MOOAC as proposed by Yu et al.~\cite{yu2023multi} as a second category method.  Nor is DEJIT~\cite{yang2020dejit,yang2021dejit} is in the second category, since it trains classifier from a differential evolution algorithm and make ranking prediction by Label/LOC.  The lesson hiding behind this is that, without the prior knowledge of the defect/LOC ratio, the performance of rank by regression is likely to be poor in the comparison with classifier-based ranking calculation.


Rao et al.~\cite{rao2021learning} proposed EA learning to rank (EALTR) and Yu et al.~\cite{yu2024improving} improve it by adding a re-ranking strategy to reduce initial false alarms. EATLR* performed better than some classifier-base methods whose IFA (initial false alarms) values are less than 10.

Yu et al.\cite{yu2023finding} analysed 34 algorithms to directly predict the ranking score includes classification, regression, pair-wise and list-wise ranking methods.  Among those algorithms with low IFA values, LTR-linear performs best under a cross-release setting, and Ranking SVM performs best under a cross-project setting.  This "low IFA value" rule excludes MunaulUP which is unsupervised and powerful in terms of Recall@20\% and Popt.

The third category is simple unsupervised methods. Yang et al.~\cite{yang2016effort} first found that simple unsupervised models could be better than supervised models (both classification and EALR models \cite{kamei2013large}).  Their unsupervised models simply sort software modules by the single attribute of defect prediction datasets.  Similarly, Liu et al.~\cite{liu_J2017ccum} used code churn to build a code churn-based unsupervised model (CCUM), which sorts software modules by lines of change.  CCUM performed better than EALR~\cite{kamei2013large}, TLEL~\cite{yang2016effort} and Yang's unsupervised models~\cite{yang2016effort}.  Further, Zhou et al.~\cite{zhou2018far} found that the simple module size model named ManualUp has a prediction performance comparable, or even superior, to most of previous cross-project defect prediction models before their work.\footnote{Note that if we consider lines of change as the LOC of commits, then CCUM is equivalent to ManualUp.
We refer to both CCUM and ManualUp as ManualUp in this paper.}

As suggested by Ni et al.~\cite{ni2022just}, in terms of Recall@20\% and Popt, Churn(ManualUp) performs better than CBS+~\cite{huang2019revisiting,ni2022just} particularly in the presence of highly skewed dataset distributions by inspecting many more modules includes non-defective ones. In other words, ManualUp finds more bugs with higher IFA scores.  It is a rather trade-off between IFA and Recall@20\%. Recent studies~\cite{yu2023multi,yu2023finding,yu2024improving} chose to improve ranking methods with low IFA, i.e. IFA<10, which find less defects than ManualUp.  Disagreeing slightly with that, we think IFA needs to be reduced whilst finding no fewer than defects than ManualUp.  So, we aim to improve classifier-based methods to find more defects with IFA less than baseline ManualUp by reducing ranking errors.

\section{Effort-aware Ranking Prediction} \label{sec:mtd}

\subsection{The Problem of Minor Chaos} \label{sec:minor_chaos}

\begin{figure}[ht]
	\includegraphics[width=1\columnwidth]{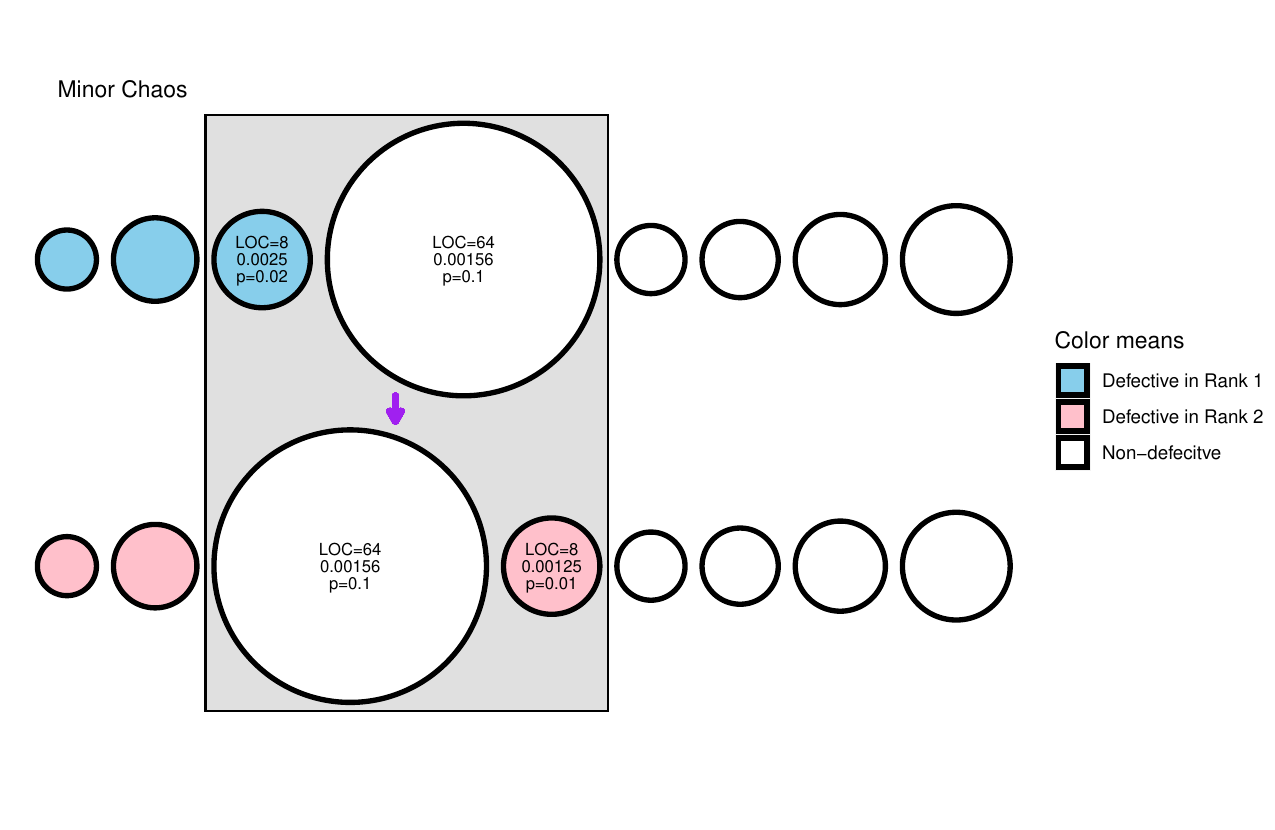}
	\caption{Example of Minor Chaos}
	\label{fig:minor}
\end{figure}

The motivation of this paper is the problem we found in predicting non-defective components for an effort-aware ranking.  We call this "Minor Chaos" since minor errors in defective probability can lead to a great difference for the Prob/LOC ratio which then leads to ranking errors.  Consider an example as shown in Fig.~\ref{fig:minor}, there are two similar rankings of defect-candidates for inspection.  The only difference is the predicted defective probability of one defect-candidate with p=0.02 in the blue upper rank and p=0.01 in the pink lower rank.\footnote{It is possible that the predicted probability can be as low as 0.02 if it is a false negative prediction i.e., a defect is wrongly predicted as defect-free.}

For classification probability, the difference between two probabilities is only 0.01, which is negligible for a classifier fitting error.  However, the ratio Prob/LOC in the upper rank is 0.0025 which is twice as much as the 0.00125 which is the ratio of the lower rank.  As a result, the rank changes and its inspection is delayed after the big white defect-candidate.  This illustrates the ranking error caused by "Minor Chaos".

\begin{figure}[ht]
	\includegraphics[width=0.7\columnwidth]{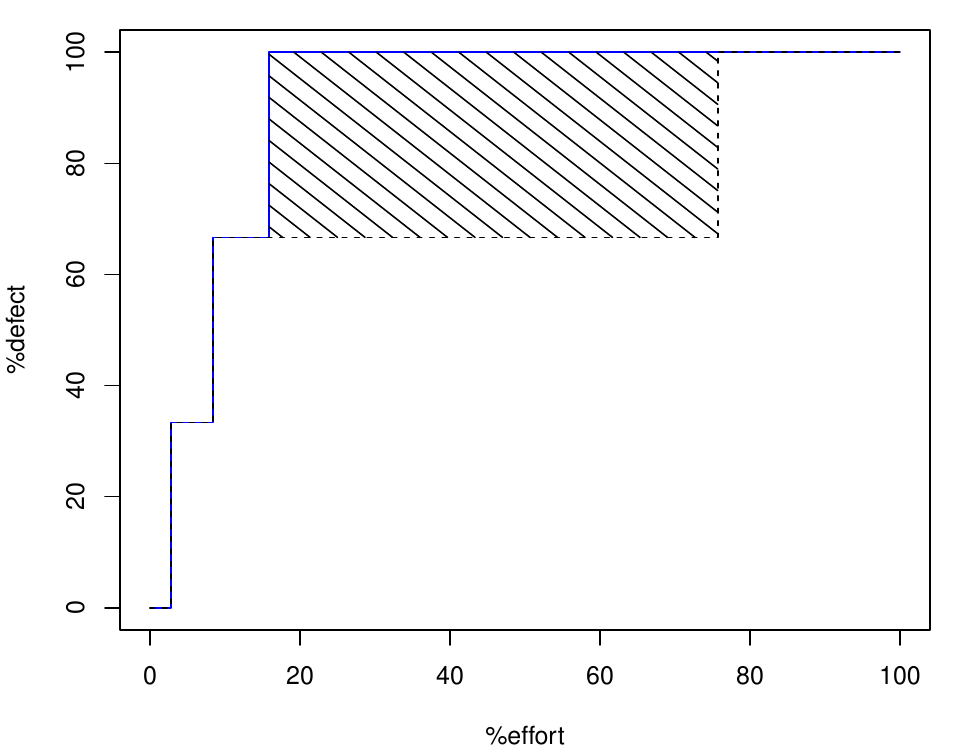}
	\caption{Minor Chaos Reduce Performance}
	\label{fig:minor2}
\end{figure}

As a result, position switching lowers the cost-effectiveness curve as shown in Fig.~\ref{fig:minor2} and the performance is harmed.
The delay of inspection reduces recall of defects with 20\% effort from 1.0 to 0.67, and lose the shadow area in the cost-effectiveness curve.
The cost of a single rank error is so significant because it requires much more effort than others to inspect the white big defect-candidate.
Unfortunately, the effort(LOC) is naturally skewed in real datasets.
So the demonstrated rank error with large cost is both possible and probable.

Furthermore, since we expect non-defective to be the majority class in software projects most predictions should be non-defective and the fitting target of defective probability will be close to zero.  Hence Minor Chaos can be easily triggered.

\subsection{Ranking Score Calculation Strategy EA-Z}


To handle Minor Chaos, we propose an effort-aware ranking score calculation strategy called EA-Z.
EA-Z setups a lower bound $\zeta$ to Prob/LOC to prevent zero-close probability prediction from classifiers.
That is why we name our method as EA-Z where Z is short for Zeta.
EA-Z calculates the effort-aware ranking score by following formula:
\[
EA_{Z}(x)=\frac{p'(x)}{LOC}
\]
where $p'(x)$ is predicted classification probability not less than $\zeta$:

\[ p'(x)=p(x)\cdot (1-\zeta)+\zeta \]

We map the original classification probability prediction $p(x)$ to new $p'(x)$.
Then the predictive interval is mapped from [0,1] to [$\zeta$,1], where $\zeta \in(0,1)$.
We prefer $\zeta $ as small as possible, when $\zeta$ is small enough, EA-Z[$\zeta$] is approximation of the defect/LOC ratio.
However at the same time, 
Lower bound $\zeta$ also has to be large enough to against "Minor Chaos".
We intuitively set $\zeta =0.05$,
Further analysis in Section~\ref{sec:zeta} shows $\zeta=0.05$ is a good choice. 

\section{Evaluation and Experimental Method} \label{sec:method}

Cross-project and cross-version experiments are conducted to compare ranking strategies with multiple classification learners.
This section gives the details of our experiments, including performance measures, data sets and classifier learners.

\subsection{Evaluation}

All effort-aware performance criteria are calculated from the CE curve in Fig.\ref{fig:curve} proposed by Arisholm.
Among several effort-aware performance criteria,
Popt\cite{mende2009revisiting,dAmbros2012evaluating,kamei2013large,yang2016effort,fu2017revisiting,chen2018multi,yu2024improving} is the most popular one.
Popt calculates the area between the curve of prediction model and the curve of optimal model. The normalised version of $P_{opt}$~\cite{yang2016effort} is
\[P_{opt}= 1- \frac{Area(optimal)-Area(m)}{Area(optimal)-Area(worst)}\]

Recall@20\%~\cite{kamei2013large,yang2016effort,fu2017revisiting,chen2018multi} is the most practical effort-aware performance criteria. 
Recall@20\% score calculates recall of defects at 20\% effort-cost\footnote{This is based on the so-called 20-80 principle.}, it calculates the proportion of the inspected actual defective modules among all the actual defective modules in the dataset.
\[ Recall@20\% = \frac{k}{K } \]
where k is the number of inspected actual defective modules and K is number of all the actual defective modules.
A higher Recall@20\% value denotes that more defective software modules could be found. 
It is also known as ACC~\cite{kamei2013large}.
This is most practical and interpretable effort-aware performance measure.

Some studies may use PofB20~\cite{yu2024improving}, which calculate the proportion of defects found with 20\% effort. If each defective module only has 1 or 0 defect, PofB20 is equal to Recall@20\%.
But not all defect prediction datasets reports the bug numbers, so we simply consider to predict whether modules are defective or not.

One more popular effort-aware performance measure in recent studies~\cite{huang2019revisiting,li2023revisiting,yu2024improving} is IFA.
IFA is the number of Initial False Alarms encountered before software testers detect the first bug.
Note unlike other performance indicators, for IFA smaller is better.

\subsection{Datasets and Data Preparation}
We conduct 61 cross-project and cross-version experiments with 72 real-world datasets in four repositories: PROMISE ~\cite{boetticher2007promise} AEEEM ~\cite{dAmbros2012evaluating}, Kamei~\cite{kamei2013large} and JavaScript datasets ~\cite{ni2022just}.  The details are summarized in Table~\ref{tbl:dataset}.

\begin{table}[ht]
\caption{Dataset Summary}
\label{tbl:dataset}
\footnotesize
\begin{tabular}{cccc}
    \toprule
    Name & \#Dataset & Type & Description \\ 
    \midrule
    PROMISE & 41 & file/class-level & code metrics \\
    AEEEM 	& 5 & class-level & code metrics and change metrics\\
    Kamei 	& 6 & commit-level & JiT metrics\\
    JavaScript 	& 20 & commit-level & JiT metrics \\
    \bottomrule
\end{tabular}
\end{table}

For the PROMISE datasets, we conducted cross-version experiments as per Yu et al.~\cite{yu2024improving}. There are 30 paired cross-version experiments on PROMISE datasets.  For the remaining datasets, we conducted cross-project experiments which require pairs of training data and testing data. To match the cross-version experiments, each dataset is used as testing data once, the training data is a datasets from the same data source with similar skewness of effect(LOC).  Skewness can be calculated as follows:
\[Skewness=\frac{1}{N} \cdot \sum_{i=1}^{n} \frac{(x_i-\bar{x})^3}{\sigma ^3} \]
where $N$ is the number of instances; $x_i$ is the effort; $\bar{x}$ is the sample mean of effort; $\sigma$ is the standard deviation.  The detail of training and testing pairs are listed in Table~\ref{tbl:cross-project}.

\begin{table}[ht]
\centering
\caption{Cross-Project Prediction Setup}
\label{tbl:cross-project}
\footnotesize
\begin{tabular}{lllr}
  \hline
  Data Source & Training Data & Testing Data & Testing Data \\ 
  & & & Skewness\\ \hline
  PROMISE & ant-1.3 & ant-1.4 & 2.08 \\ 
  PROMISE & ant-1.4 & ant-1.5 & 3.95 \\ 
  PROMISE & ant-1.5 & ant-1.6 & 3.44 \\ 
  PROMISE & ant-1.6 & ant-1.7 & 3.84 \\ 
  PROMISE & camel-1.0 & camel-1.2 & 2.90 \\ 
  PROMISE & camel-1.2 & camel-1.4 & 3.74 \\ 
  PROMISE & camel-1.4 & camel-1.6 & 4.12 \\ 
  PROMISE & ivy-1.1 & ivy-1.4 & 8.78 \\ 
  PROMISE & ivy-1.4 & ivy-2.0 & 3.32 \\ 
  PROMISE & jedit-3.2 & jedit-4.0 & 11.21 \\ 
  PROMISE & jedit-4.0 & jedit-4.1 & 11.12 \\ 
  PROMISE & jedit-4.1 & jedit-4.2 & 6.88 \\ 
  PROMISE & jedit-4.2 & jedit-4.3 & 7.71 \\ 
  PROMISE & log4j-1.0 & log4j-1.1 & 1.57 \\ 
  PROMISE & log4j-1.1 & log4j-1.2 & 4.57 \\ 
  PROMISE & lucene-2.0 & lucene-2.2 & 4.95 \\ 
  PROMISE & lucene-2.2 & lucene-2.4 & 7.59 \\ 
  PROMISE & poi-1.5 & poi-2.0 & 10.37 \\ 
  PROMISE & poi-2.0 & poi-2.5 & 9.33 \\ 
  PROMISE & poi-2.5 & poi-3.0 & 9.42 \\ 
  PROMISE & synapse-1.0 & synapse-1.1 & 1.82 \\ 
  PROMISE & synapse-1.1 & synapse-1.2 & 1.98 \\ 
  PROMISE & velocity-1.4 & velocity-1.5 & 10.10 \\ 
  PROMISE & velocity-1.5 & velocity-1.6 & 10.54 \\ 
  PROMISE & xalan-2.4 & xalan-2.5 & 3.19 \\ 
  PROMISE & xalan-2.5 & xalan-2.6 & 3.04 \\ 
  PROMISE & xalan-2.6 & xalan-2.7 & 3.04 \\ 
  PROMISE & xerces-init & xerces-1.2 & 5.19 \\ 
  PROMISE & xerces-1.2 & xerces-1.3 & 5.61 \\ 
  PROMISE & xerces-1.3 & xerces-1.4 & 5.05 \\ 
  AEEEM & Equinox-AEEEM & PDE-AEEEM & 3.24 \\ 
  AEEEM & PDE-AEEEM & Equinox-AEEEM & 3.79 \\ 
  AEEEM & Equinox-AEEEM & Lucene-AEEEM & 6.26 \\ 
  AEEEM & Lucene-AEEEM & JDT-AEEEM & 7.50 \\ 
  AEEEM & JDT-AEEEM & Mylyn-AEEEM & 20.54 \\ 
  Kamei & columba & bugzilla & 19.64 \\ 
  Kamei & bugzilla & columba & 36.74 \\ 
  Kamei & columba & mozilla & 47.55 \\ 
  Kamei & mozilla & postgres & 81.04 \\ 
  Kamei & postgres & jdt & 100.85 \\ 
  Kamei & jdt & platform & 141.30 \\ 
  JavaScript & anime & Chart.js & 7.23 \\ 
  JavaScript & Chart.js & anime & 9.67 \\ 
  JavaScript & anime & jquery & 9.89 \\ 
  JavaScript & jquery & parcel & 10.86 \\ 
  JavaScript & parcel & axios & 11.11 \\ 
  JavaScript & axios & express & 12.38 \\ 
  JavaScript & express & vue & 13.35 \\ 
  JavaScript & vue & yarn & 13.37 \\ 
  JavaScript & yarn & hyper & 15.80 \\ 
  JavaScript & hyper & three.js & 16.63 \\ 
  JavaScript & three.js & material-ui & 21.61 \\ 
  JavaScript & material-ui & Ghost & 24.03 \\ 
  JavaScript & Ghost & lodash & 27.25 \\ 
  JavaScript & lodash & babel & 27.97 \\ 
  JavaScript & babel & react & 29.04 \\ 
  JavaScript & react & moment & 29.41 \\ 
  JavaScript & moment & serverless & 31.52 \\ 
  JavaScript & serverless & webpack & 35.43 \\ 
  JavaScript & webpack & pdf.js & 66.72 \\ 
  JavaScript & pdf.js & meteor & 82.76 \\ 
   \hline
\end{tabular}
\end{table}

From Table~\ref{tbl:cross-project} we observe that the skewness of the datasets range from a minimum of 1.57 to a maximum of 141.3 and a median of 9.67.  Thus, the real-world data sets are all strongly right-skewed, which exacerbate the problem of Minor Chaos as per the example in Section~\ref{sec:minor_chaos}.

In our experiment, we do the following data pre-processing:
\begin{itemize}
\item Apply logarithmic transformation to JiT attributes following Yang et al.~\cite{yang2016effort}. Since the input attributes are also highly skewed, we do the same pre-processing to alleviate this effect as well.
\item Remove zero LOC or churn instances from the datasets. It is possible to submit an empty commit when no changes have occurred. However, empty commits and empty files are considered as invalid data and removed.
\end{itemize}

\subsection{Learners and Classifier Settings} \label{sec:algorithm}

All parameter and settings must be specified in order to run experiments. Since our study EA methods are based on classifiers, the settings are mainly related to the classification learners.  The settings of learners includes the choice of classification algorithms and parameters in algorithms. 
For classification algorithms, we include various types of techniques: function~(Logistic Regression), rule~(JRip), bayes~(Naive-Bayes), decision Tree~(C50, cart, Random-Forest), Lazy (K-nearest Neighbour) and ensemble learning (Under-Bagging and RUSBoost)\footnote{Imbalanced ensemble methods UnderBagging and RUSBoost are from R package "embc"}. These are summarised in Table~\ref{tbl:learner}.

\begin{table}[ht]
	\caption{Overview of Classification Learners}
	\label{tbl:learner}
	\footnotesize
	\begin{tabular}{lll}
		\hline
		Category & Algorithm & Abbreviation \\
		\hline
		Function & Logistic Regression & LR \\
				 & Support Vector Machine & SVM \\
		\hline
		Lazy & K-Nearest Neighbour & IBk \\
		\hline
		Rule & Propositional rule & JRip \\
		\hline
		Bayes& Naive Bayes & NB \\
		\hline
		   Tree & Decision Tree & C50 \\
			 & Classification and Regression Trees  & cart  \\		   
			 & Random Forest        & RF  \\
		\hline
		Ensemble & UnderBagging & UBag-c50,UBag-rf \\
                    & & UBag-cart,UBag-svm \\
				 & RUSBoost     & UBst-c50,UBst-rf \\
                     & & UBst-cart,UBst-svm \\
		\hline
	\end{tabular}
\end{table}

In total, there are 16 learners. We include imbalanced ensemble learning because the distribution of labels for defect prediction datasets are typically imbalanced \cite{song2018imb}, and an effective way to handle it is via imbalanced ensemble learning~\cite{song2018imb}, i.e., under-sampling ensemble methods (UBag = under-sampling + bagging and UBst = under-sampling + boosting).

The parameter settings of the learners are as follows:
\begin{itemize}
    \item For subsampling in UBag and UBst, set $ir=1$ which means equal number of positive and negative in training data;
    \item For k-Nearest Neighbor (IBk), set $k=8$ in as same as FSE16~\cite{yang2016effort};
    \item RandomForest tree number is 200 alone, but 50 in ensemble methods e.g., UBag or UBst.
    \item The kernel method of SVM is set as "radial".
    \item We set the classification threshold as 0.5 which is the same as the ranking method CBS+ from Huang et al.~\cite{huang2019revisiting} and Li et al.~\cite{li2023revisiting}.
\end{itemize} 

\section{Experimental results} \label{sec:results}

In this section, we first examine the five EA ranking score calculation strategies: (1) Prob; (2) Label/LOC; (3) CBS+; (4) Prob/LOC; (5) EA-Z.
Then we analyse the impact of the different classification learners and the settings of the parameter $\zeta$ in our method EA-Z.

\subsection{Ranking Strategy Comparison}

\begin{figure}[ht]
	\includegraphics[width=0.9\columnwidth]{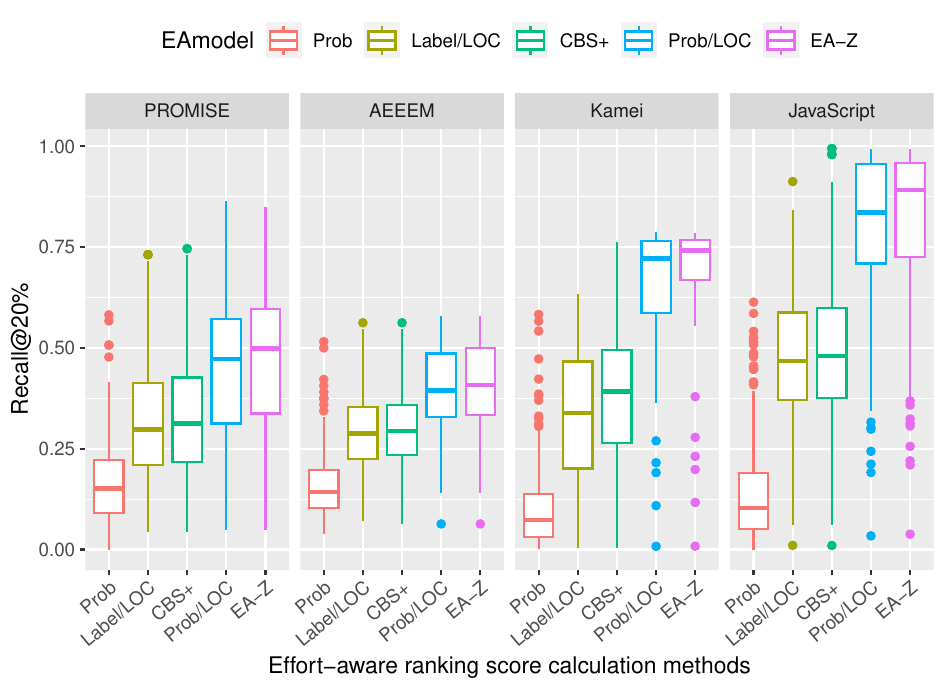}
	\caption{Recall@20\% performance of 5 ranking strategies}
	\label{fig:rankacc}	
	
\end{figure}

\begin{figure}[ht]
	\includegraphics[width=0.9\columnwidth]{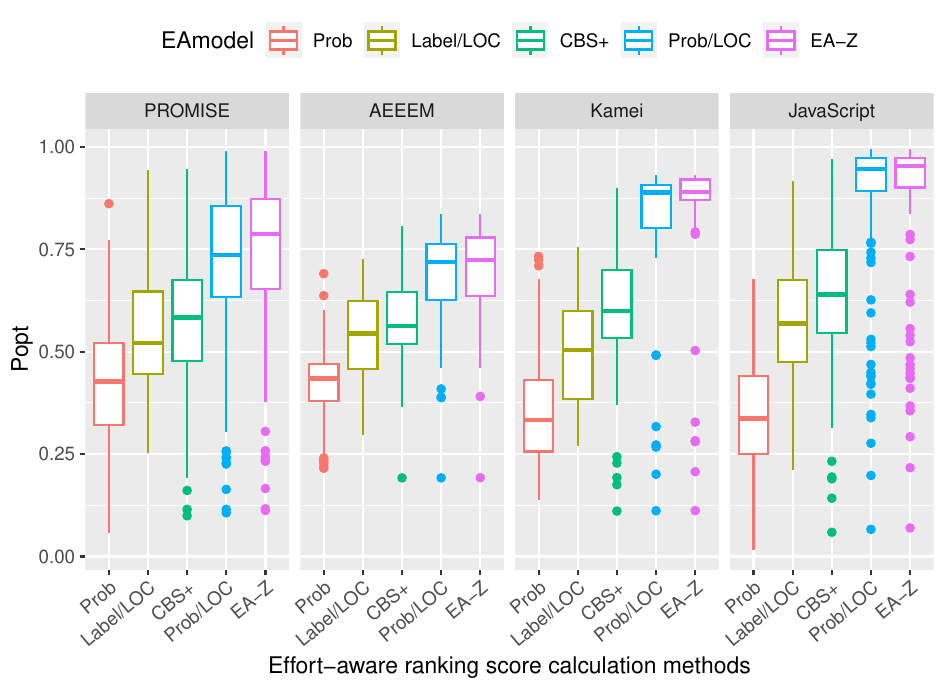}
	\caption{Popt performance of 5 ranking strategies}
	\label{fig:rankpopt}	
	
\end{figure}

\begin{figure}[ht]
	\includegraphics[width=0.9\columnwidth]{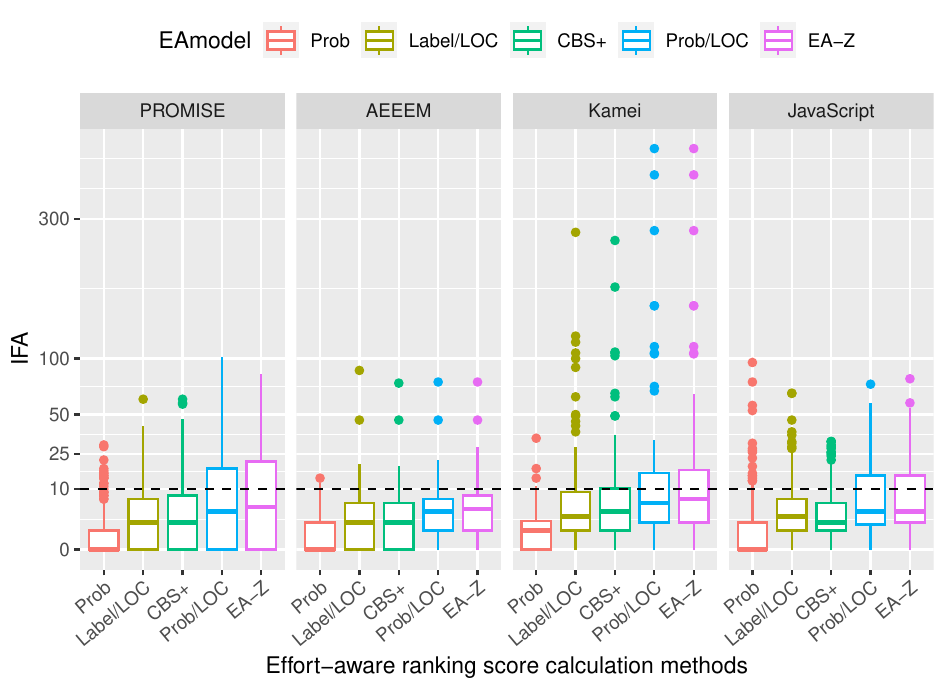}
	\caption{IFA performance of 5 ranking strategies}
	\label{fig:rankIFA}	
	
\end{figure}

To present a comprehensive view of the five ranking strategies, 
we apply each strategy with 16 classification learners and conduct 61 cross-project and cross-version experiments with 72 real-world datasets as shown in Table~\ref{tbl:cross-project}.  This means that there are 16*61=976 results for each ranking strategy.  The results are presented in box-plots in Figs.~\ref{fig:rankacc}, \ref{fig:rankpopt} and \ref{fig:rankIFA} in terms of three performance indicators respectively.

The first performance indicator is Recall@20\% which shows the recall of defects by inspecting only 20\% LOC. The results of Recall@20\% shown in Fig.~\ref{fig:rankacc} are grouped by the four data sources: PROMISE, AEEEM, Kamei and JavaScript.  In each group, there are five boxes with whiskers to visualize the distribution of the performance scores of five ranking strategies: Prob, Label/LOC, CBS+, Prob/LOC and EA-Z.
The upper edge of a box is the upper quartile, and the lower edge of a box is the lower quartile. The whiskers extend from the box to show the variability outside the upper and lower quartiles, and the middle line in a box denotes the median.

From Fig.~\ref{fig:rankacc} we observe that for all four groups of datasets, the medians from low to high are Prob, Label/LOC, CBS+, Prob/LOC and EA-Z.  That means that typically, EA-Z achieves the highest Recall@20\% performance among the five ranking strategies.

For detailed comparisons, we calculate more statistics in Table~\ref{tbl:g_recall} including mean, win/draw/loss count and the Wilcoxon signed rank tests.  The Win/Draw/Loss record uses paired comparisons to represent the number of times EA-Z performs better than other ranking strategies in this paper.  The Wilcoxon signed rank test is a non-parametric statistical test used to compare EA-Z with the other four ranking strategies to determine whether EA-Z is statistically superior.  Note that the p-values of these test are adjusted using the False Discovery Rate correction~\cite{benjamini2001control} to counteract the problem of multiple tests. In addition, we also calculate the effect size $r$ for the Wilcoxon signed-rank test~\cite{tomczak2014need}. A simplistic interpretation of $r$ is that, effect size is considered trivial for $|r|<0.1$, small for $0.1<|r|<0.3$, moderate for $0.3<|r|<0.5$, large for $|r|>0.5$.

\begin{table}[ht]
    \centering
    \caption{Average Recall@20\% and Comparison}
    \label{tbl:g_recall}
    \footnotesize
    \begin{tabular}{rlllll}
        \hline
        Method & EA-Z & Prob & Label/LOC & CBS+ & Prob/LOC \\  \hline
        Average & 0.605 & 0.153 & 0.37 & 0.389 & 0.587 \\ 
        W/D/L & - & 960/7/9 & 888/31/57 & 877/53/46 & 512/293/171 \\ 
        P-value & - & <0.001 & <0.001 & <0.001 & <0.001 \\
        Effect Size & - & 0.865 & 0.835 & 0.835 & 0.473 \\
        Interpretation & - & large & large & large & moderate \\
        \hline
    \end{tabular}
\end{table}

From Table~\ref{tbl:g_recall} we observe that the average score of EA-Z is 0.605 for Recall20\%.  That means testing on 61 datasets with 16 learners, the ranking strategy EA-Z can find 60.5\% defects by inspecting 20\% LOC.  It is the highest Recall20\% score among the five ranking strategies showing that EA-Z can find about 20\% more defects than CBS+.  In paired comparisons, EA-Z is better 960, 888, 877 and 512 times than Prob, Label/LOC, CBS+ and Prob/LOC respectively.  In each case this represents more than half of the 976 comparisons.  The p-values are all less than 0.001 after correction which offers evidential support to the idea of meaningful differences in performance.  The effect size is moderate for comparing EA-Z with Prob/LOC, and otherwise large.

Similar to Fig.~\ref{fig:rankacc}, Fig.~\ref{fig:rankpopt} presents the results of the second performance indicator Popt.  From Fig.~\ref{fig:rankpopt} we observe medians, from low to high, are for Prob, Label/LOC, CBS+, Prob/LOC and EA-Z for the PROMISE, AEEEM and JavaScript datasets.  As for the Kamei datasets, the median of EA-Z is as high as Prob/LOC but the box is smaller suggesting less variance.  Overall EA-Z has higher a Popt score distribution.  To corroborate this, we provide additional summary statistics shown in Table~\ref{tbl:g_popt}.

\begin{table}[ht]
	\caption{Average Popt and Comparison}
    \label{tbl:g_popt}
	\footnotesize
	\centering
	\begin{tabular}{rlllll}
	  \hline
	  Method  & EA-Z & Prob & Label/LOC & CBS+ & Prob/LOC \\ \hline
	  Average & 0.813 & 0.389 & 0.549 & 0.602 & 0.791 \\ 
	  W/D/L & - & 953/0/23 & 936/0/40 & 949/6/21 & 727/6/243 \\ 
	  P-value & - & <0.001 & <0.001 & <0.001 & <0.001 \\ 
   	Effect Size & - & 0.861 & 0.841 & 0.862 & 0.593 \\ 
	  Interpretation & - & large & large & large & large \\ 
	\hline
\end{tabular}
\end{table}

From Table~\ref{tbl:g_popt} we observe that the mean score of EA-Z is 0.813 for Popt which is best for the five columns.  In a paired comparison, EA-Z is better for 953, 936, 949 and 727 times than the alternative four ranking strategies respectively.  In each case this is greater than half of the 976 comparisons and all p-values are less than 0.001 after correction.
This suggests that EA-Z is meaningfully better than the other ranking strategies in terms of Popt.  Furthermore, all of the effect sizes are large.

The third performance indicator is initial false alarms (IFA) before the first detection of defects.  That means IFA is better to be lower.  However, there is something of a trade-off in that a prediction rank can achieve higher recall by inspecting more software modules with more false alarms.  To quantify the trade-off, we present the results of IFA shown in Fig.~\ref{fig:rankIFA}.

To improve the readability of Fig.~\ref{fig:rankIFA}, we perform square root transformation was on the y-axis scale, because there are some extreme outliers which would make the range of the Y-axis too wide compared with the size of boxes in the box-plot. The dotted line we added is where IFA equals to 10 since previous researchers have also focused on EA methods with IFA<10~\cite{huang2019revisiting,ni2022just,yu2023finding,yu2024improving}.

From Fig.~\ref{fig:rankIFA}, we observe the order of medians is as same as other two performance indicators. So it suggests the trade-off in EA ranking such that the better the predicted ranking found, the higher the IFA. Although no strategy is entirely under the dotted line, but the major part of each box does achieve this performance threshold.  In particular, note that 64.8\% of the predicted ranks of EA-Z are no more than 10 in terms of IFA.  More statistics are provided in Table~\ref{tbl:g_IFA}.

\begin{table}[ht]
    \caption{Average IFA and Comparison}
    \label{tbl:g_IFA}
	\footnotesize
    \centering
    \begin{tabular}{rlllll}
        \hline
        Method  & EA-Z & Prob & Label/LOC & CBS+ & Prob/LOC \\   \hline
        Average & 14.198 & 1.853 & 6.337 & 6.497 & 13.25 \\
        W/D/L & - & 109/173/694 & 216/215/545 & 155/322/499 & 103/730/143 \\ 
        \hline
    \end{tabular}
\end{table}

The mean IFA of EA-Z is 14.198.  Although it exceeds the 10 threshold, the recall score is increased from 0.389 (CBS+) to 0.605 at the cost of about 8 more false alarms.  We recommend it as a more efficient option, in order to find about 20\% more defects with the expectation of an IFA of less than 15.

In summary among five ranking strategies, EA-Z is best in median scores of Recall@20\% and Popt. Overall EA-Z is the best ranking strategy given the expectation that IFA is less than 15.

\subsection{Learner Comparison and Baseline}

In this subsection we study the ranking methods of EA-Z with a specific classifier.  In order to compare with previous works~\cite{yu2024improving,ni2022just,li2023revisiting,yu2023multi}, we add three popular baseline methods: ManualUp~\cite{zhou2018far}, CBS+(LR)~\cite{huang2019revisiting} and CBS+(RF)~\cite{li2023revisiting}.  In total there are 16+3=19 methods for comparison, i.e., EA-Z with 16 learners and 3 baseline methods. Details of them can be found in Section~\ref{sec:RelWork}.

\begin{figure}[ht]
	\includegraphics[width=0.9\columnwidth]{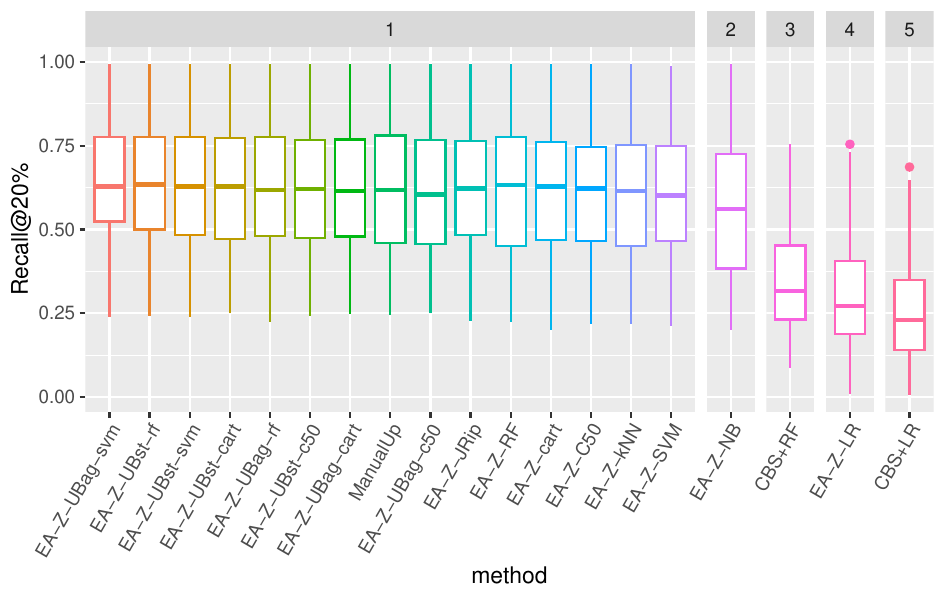}
	\caption{Recall@20\% performance of EA-Z and baselines}
	\label{fig:approach_acc}	
	
\end{figure}

\begin{figure}[ht]
	\includegraphics[width=1\columnwidth]{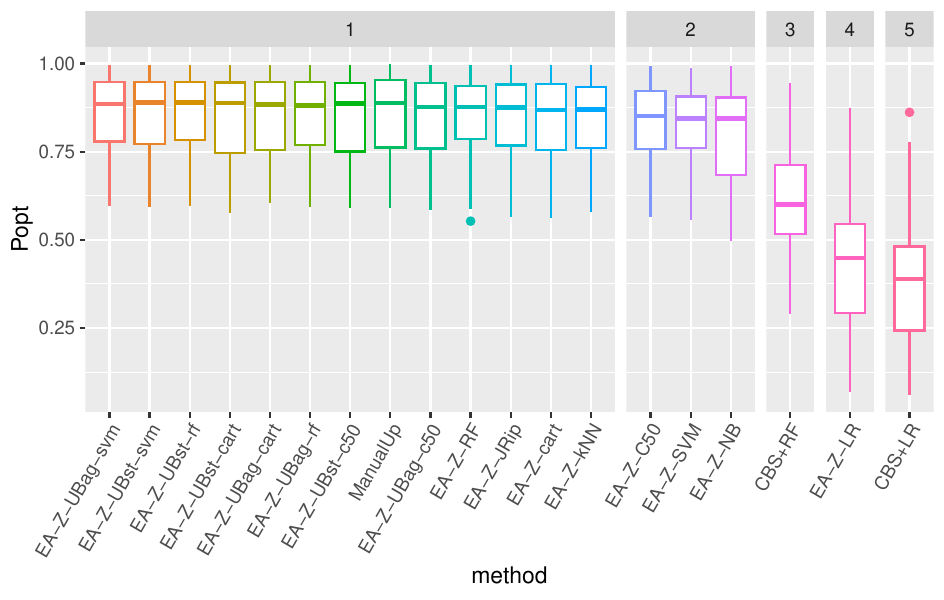}
	\caption{Popt performance of EA-Z and baselines}
	\label{fig:approach_popt}	
	
\end{figure}

\begin{figure}[ht]
	\includegraphics[width=0.9\columnwidth]{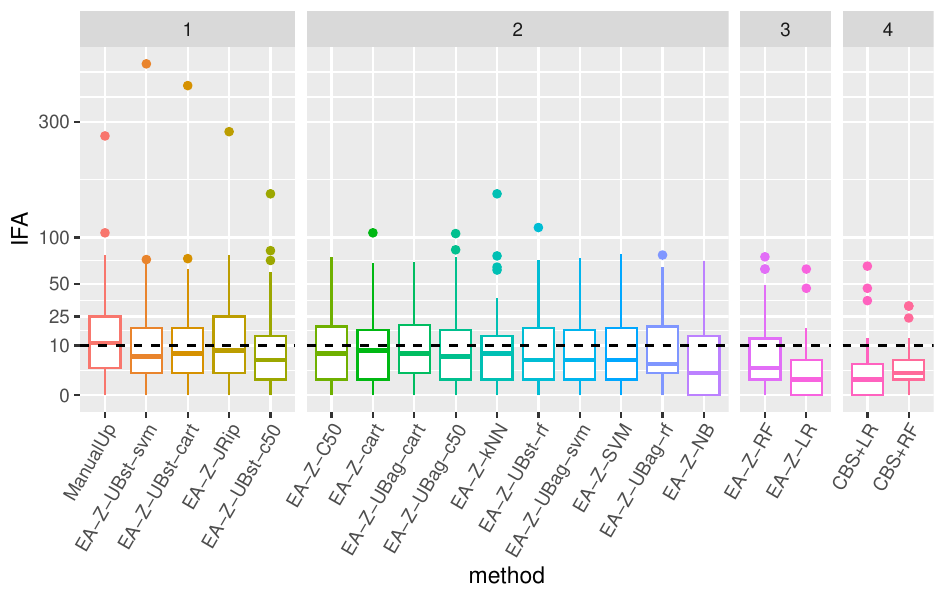}
	\caption{IFA performance of EA-Z and baselines}
	\label{fig:approach_IFA}	
	
\end{figure}

The results of these ranking methods are shown in Fig.~\ref{fig:approach_acc}, Fig.~\ref{fig:approach_popt} and Fig.~\ref{fig:approach_IFA} in terms of Recall@20\%, Popt and IFA respectively.  The results of these ranking methods are grouped by numbers which are the group number from the Scott-Knott Effect Size Difference (ESD)~\cite{tantithamthavorn2018impact} which is a statistical test for comparing multiple methods on multiple datasets.  It clusters the set of performance averages (e.g., means) into statistically distinct groups with non-negligible difference. We used the Non-Parametric version of Scott-Knott ESD test by R Package ScottKnottESD~\footnote{https://github.com/klainfo/ScottKnottESD}.

Fig.~\ref{fig:approach_acc} presents the boxplot of 19 ranking methods, which are sorted by mean Recall@20\% score from high to low. From Fig.~\ref{fig:approach_acc} we can observe that there are 5 groups numbered from 1 to 5.  EA-Z with 14 different learners and ManualUp are in the 1st group, and then EA-Z-NB, CBS+RF, ZA-Z-LR, CBS+LR are in the second, third, fourth and fifth groups respectively.  The performance of methods in the first group is statistically better than those of the second/subsequent groups.  That means the 14 EA-Z methods in the first group are statistically higher than the baseline CBS+LR and CBS+RF, and similarly good as ManualUp.  Further, seven EA-Z methods are sorted before ManualUp, which means the average mean of them is higher than the mean of ManualUp.  Note that the learners of the seven EA-Z methods are all imbalanced ensemble learning methods, and the top two are UBag-svm and UBst-rf.

Fig.~\ref{fig:approach_popt} presents the boxplot sorted by mean Popt score from high to low.  From Fig.~\ref{fig:approach_popt} we can observe that EA-Z with 12 different learners and ManualUp are in Group 1, which is statistically higher than Group 2, Group 3 (baseline CBS+RF), Group 4 and Group 5 (baseline CBS+LR).
Seven EA-Z methods are sorted before ManualUp again as same as on Recall@20\% score. The learners of the seven EA-Z methods are all imbalanced ensemble learning methods: UBag-svm, UBst-svm, UBst-rf, UBst-cart, UBag-cart, UBag-rf and UBst-c50.  It is unsurprising that all the top learners are imbalanced ensemble learning methods since defect prediction datasets are imbalance in class label~\cite{song2018imb}.

Fig.~\ref{fig:approach_IFA} presents the boxplot sorted by average IFA from high to low.  The dotted line we added is the line that IFA equals to 10, which is the suggested line of previous studies~\cite{huang2019revisiting,ni2022just,yu2023finding,yu2024improving}. From Fig.~\ref{fig:approach_IFA} we observe that only five methods including ManualUp are in the top group (Group 1).
Among the top seven EA-Z methods in previous analysis, EA-Z-UBag-svm, EA-Z-UBst-rf, EA-Z-UBag-cart and EA-Z-UBag-rf are in Group 2.  That means the average IFA scores of these four methods are statistically lower than the baseline ManualUp while they find no fewer defects than ManualUp. If low IFA is preferred, EA-Z-RF and EA-Z-LR in Group 3, and CBS+LR and CBS+RF in Group 4 have further lower IFA.  However, it is something of a trade-off between the recall of defects and IFA.

To quantify the trade-off between Recall@20\% and IFA, we summarize the trade-off with numbers in Table~\ref{tbl:trade4}.  If IFA is strictly restricted to less than 10, then CBS+ is the choice, however, CBS+ may lose about 0.27 in Recal@20\% score when compared to ManualUp.  If IFA is allowed to be 50\% IFA of ManualUp, then EA-Z-RF can find as many defects as ManualUp.  If IFA less than 15 is acceptable, then EA-Z-UBag-svm can find 0.8\% defective modules more than ManualUp.

\begin{table}[ht]
    \caption{Trade-off between Recall@20\% and IFA}
    \label{tbl:trade4}
    \small
    \centering
    \begin{tabular}{rllll}
        \hline
        Method & ManualUp & CBS+(RF) & EA-Z-RF & EA-Z-UBag-svm \\ \hline
        Recall20 & 0.629 & 0.355 & 0.626 & 0.637 \\ 
        WDL & - & 7/1/53 & 29/5/27 & 35/10/16 \\ 
        IFA & 21.213 &  4.049 & 10.623 & 13.820 \\ 
        \hline
    \end{tabular}
\end{table}

Finally, given the expectation that IFA is less than 15, we recommend EA-Z-UBst-rf and EA-Z-UBag-svm which achieve the top performance in terms of Recall@20\% and Popt and have fewer initial false alarms than ManualUp.

\subsection{Additional analysis for Zeta} \label{sec:zeta}

\begin{figure}[ht]
	\includegraphics[width=0.9\columnwidth]{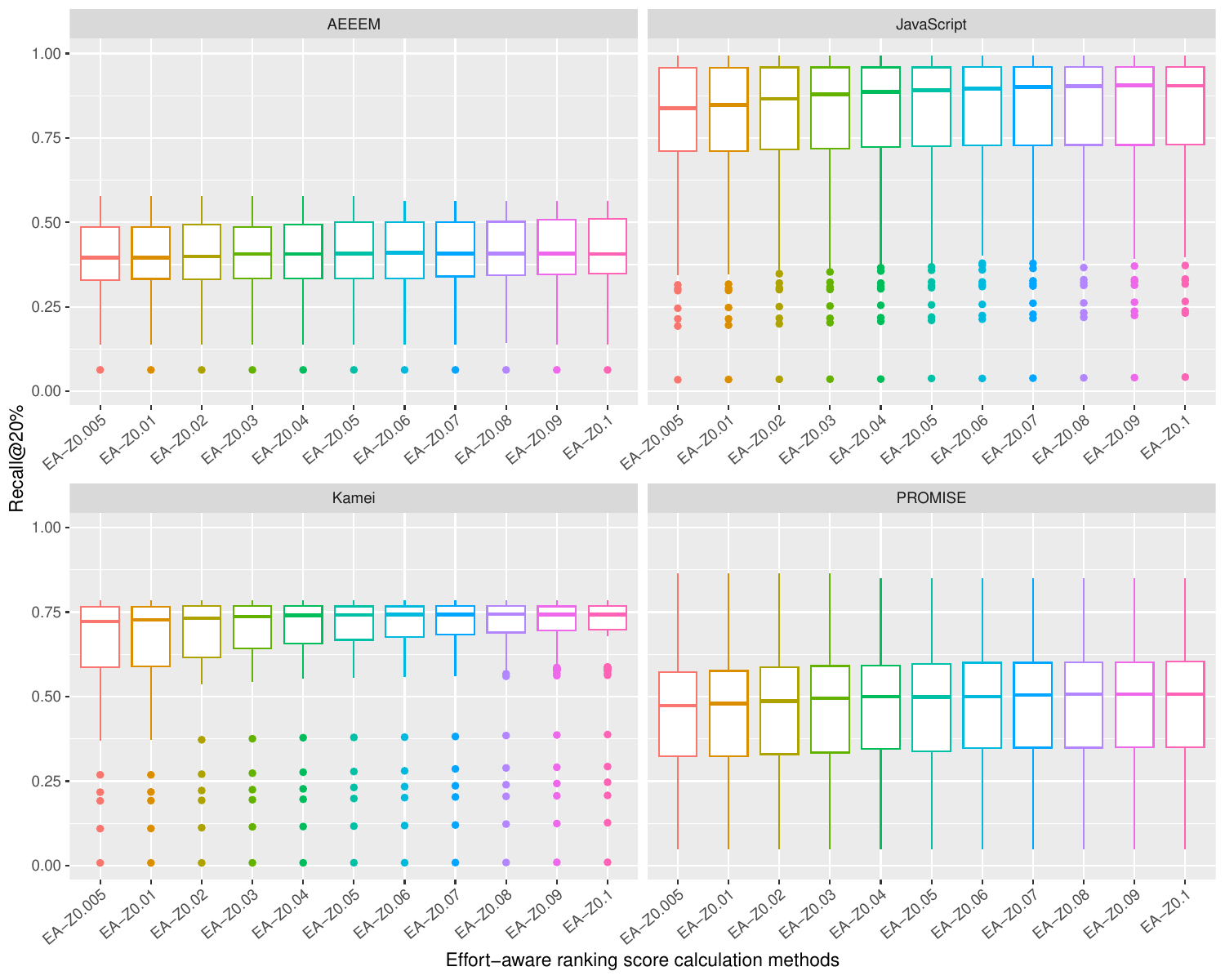}
	\caption{Recall@20\% of EA-Z with different $\zeta$}
	\label{fig:zeta_acc}	
	
\end{figure}

\begin{figure}[ht]
	\includegraphics[width=0.9\columnwidth]{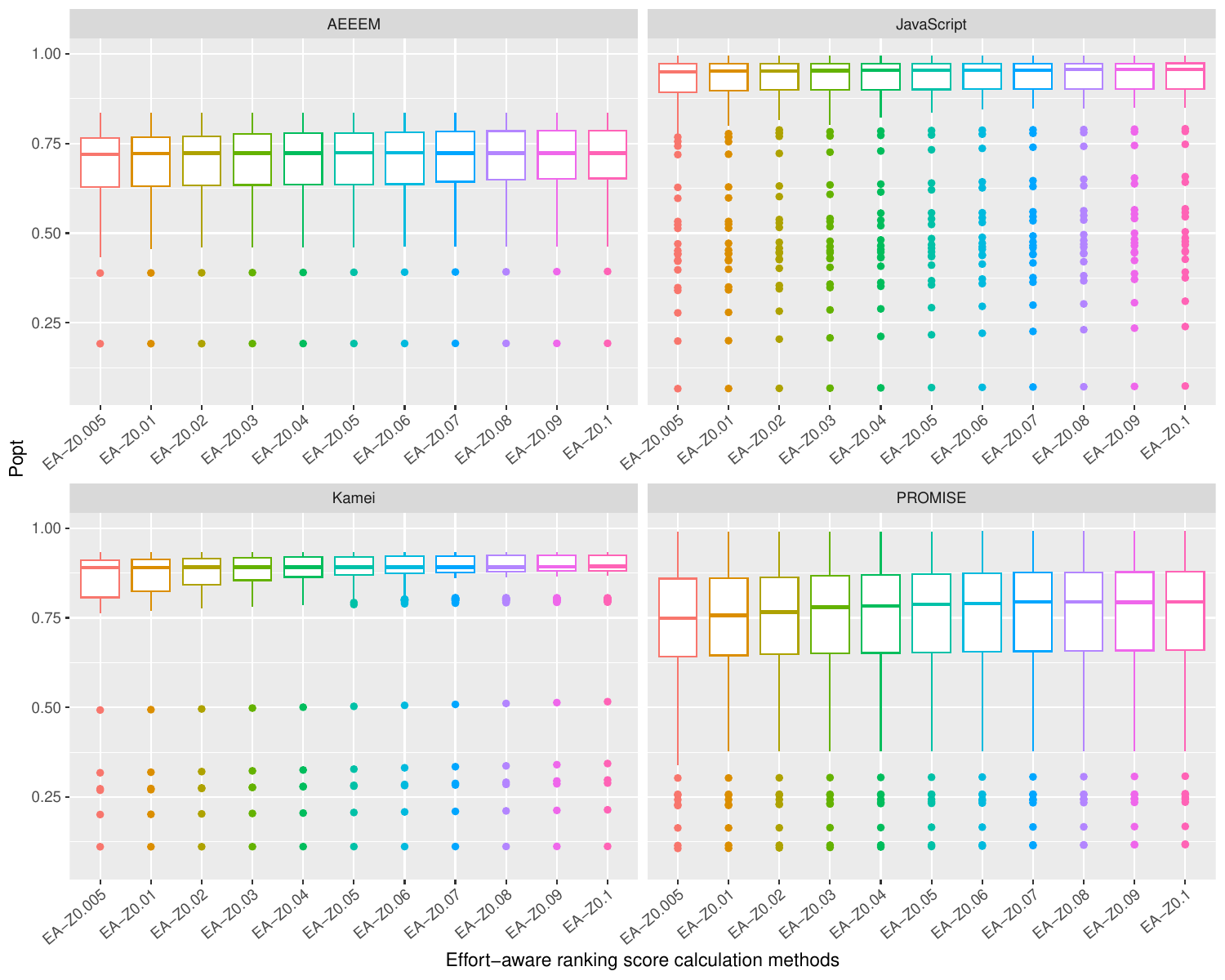}
	\caption{Popt of EA-Z with different $\zeta$}
	\label{fig:zeta_popt}	

\end{figure}

\begin{figure}[ht]
	\includegraphics[width=0.9\columnwidth]{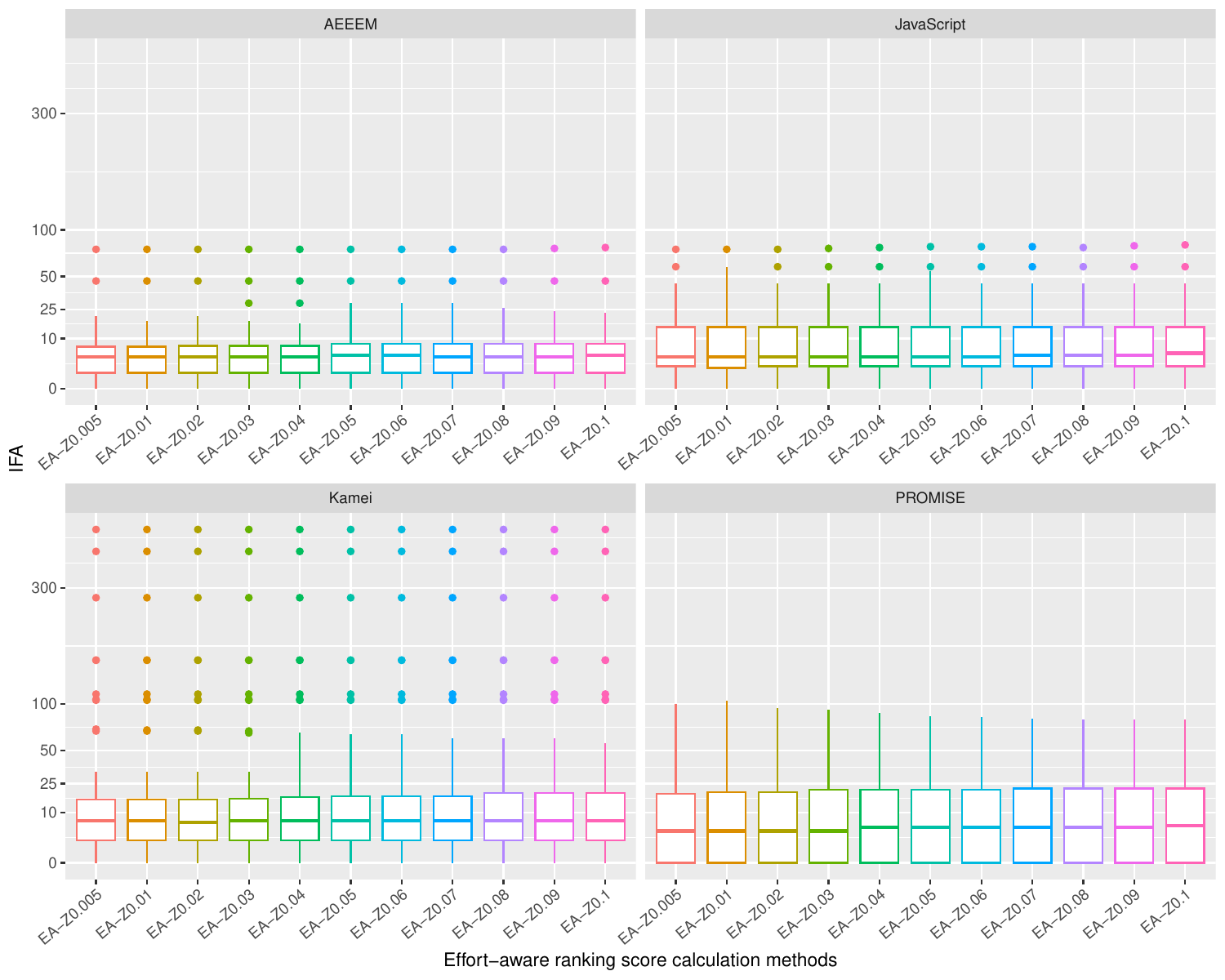}
	\caption{IFA of EA-Z with different $\zeta$}
	\label{fig:zeta_IFA}	
	
\end{figure}

In previous comparisons we set $\zeta $ in EA-Z as 0.05 since intuitively this seemed reasonable.  We prefer $\zeta $ as small as possible to approximate the defect/LOC ratio.  However at the same time, the lower bound of $\zeta$ also has to be large enough to counteract the problem of "Minor Chaos".  In this section, we study the how EA-Z performs with $\zeta $ ranging from 0.005 to 0.1.  The box-plots of three performance indicators are shown in Fig.~\ref{fig:zeta_acc}, Fig.~\ref{fig:zeta_popt} and Fig.~\ref{fig:zeta_IFA} grouped by data sources. 

From Fig.~\ref{fig:zeta_acc} we can observe that, overall on Recall@20\%, the boxes are similar except for those in Group Kamei. The main difference is that the median line rises as $\zeta$ increases from 0.005 to 0.05. In Group Kamei, the lower quantile increases as $\zeta$ increases. When $\zeta$ is greater than 0.05, though some median lines can be higher, the difference is small. Since we prefer smaller $\zeta$, 0.05 is a good choice.

Fig.~\ref{fig:zeta_popt} presents the results on Popt showing that there is very tiny difference in Group AEEEM and Group JavaScript. In Group Kamei, similar to Recall@20\%, the lower quantile increases with $\zeta$. In Group PROMISE, the median lines rise with $\zeta$.  When $\zeta$ is greater than 0.05, the difference is small on Popt as well.

From Fig.~\ref{fig:zeta_IFA} we can observe very little difference between all four groups.  In summary, setting $\zeta $ as 0.05 is a good choice for EA-Z since greater values $\zeta$ would results in very little improvement.

\section{Threats to Validity}

We conducted 61 cross-project and cross-version experiments with 72 real world datasets used in primary effort-aware software defect prediction studies~\cite{yu2024improving,li2023revisiting,huang2019revisiting,ni2022just} and benchmark software defect prediction datasets~\cite{dAmbros2012evaluating}.  The 72 datasets include, not only commit-level (just-in-time) defect datasets, but also file/class-level (traditional) datasets, and have varying numbers of modules, defective ratios and skewness for effort(LOC), which contributes to a certain generalizing capability.  The number of datasets is more than most EA software defect prediction studies,
and all datasets are publicly available online. Nevertheless, we can not be certain that EA-Z performs similarly for other datasets, so we provide experimental details and share datasets and our codes~\footnote{https://zenodo.org/records/10846586}.

Although existing cross-project studies have some techniques to select training data~\cite{ni2020revisiting}, we are less aware of the effectiveness of training data selection. 
Effectively selecting training data may enhance the training of classifiers and improve the performance of classifier-based methods.  So our initial experiments merely illustrate the basic potential or minimum capability of our method EA-Z.
We apply a simple rule to choose training data sets: use data sets with similar skewness of effort(LOC). Our experiment can be easily replicated according to Table~\ref{tbl:cross-project}.

Employing different learning algorithms may lead to variations in performance score.  The classifier learners used in this paper are all from R packages which are freely available.  Among the 16 learners used in this paper, half of them are classic machine learning algorithms.  The remainder are imbalanced ensemble learners from R package "ebmc", which is recommended for imbalanced datasets as per defect prediction datasets~\cite{song2018imb}.  The broad range of learners contribute to some generalizing capability too.  

Another potential concern is the accuracy of the data. Most notably the labelling of the software components as defective or otherwise. It is entirely possible that some of the data used is erroneous~\cite{liu2022inconsistent}, however, we note that we use many data sets and each data set comprises many instances.  Consequently, we would hope that the impact of data errors will be limited on our overall analysis.

\section{Conclusions}\label{sec:conclusion}

Existing classifier-based effort-aware prediction methods make EA prediction by adding weights to the classification prediction but they tend to make limited consideration of ranking errors.

In this paper, we have viewed EA defect prediction as a sorting problem, hence our focus errors in predicted rank as ranking errors.  We have identified a problem we name "Minor Chaos" in the ranking strategy Prob/LOC which can cause non-trivial ranking errors. It occurs when the predicted defective probability is near-zero.  Unfortunately --- at least in terms of predicting if not terms of the overall quality of the software --- non-defective instances are almost always in the majority for software projects. So most predictions should be non-defective so that Minor Chaos can be easily triggered.

To counteract Minor Chaos, we have proposed an EA ranking score calculation strategy EA-Z.  EA-Z handles "Minor Chaos" by setting a boundary $\zeta$ to prevent near-zero prediction.  We have then compared EA-Z with four other EA ranking strategies with 16 classification learners.  We conducted 61 cross-version and cross-project experiments on 72 real datasets.  Though the within-project results are more favorable to supervised methods, in practice only cross-project results are really important to relevant to software engineers.  Among the five ranking strategies, EA-Z achieves best average scores of Recall@20\% and Popt given the expectation that initial false alarms (IFA) are less than 15.  Our analysis supports the effectiveness of reducing ranking errors.

When assessing EA-Z with specific classifiers, e.g.,
EA-Z-UBag-svm, EA-Z-UBst-rf, EA-Z-UBag-cart and EA-Z-UBag-rf, it achieve better scores than the three baselines, including ManualUp. They also have fewer initial false alarms than ManualUp.  The four top learners all use imbalanced ensemble learning, which is not surprising since defect prediction datasets frequently have imbalanced class distributions. Therefore we recommend EA-Z with imbalanced ensemble learners.

Finally, we have quantified the trade-off between the recall of defects and initial false alarms. We hope this proves helpful for both software defect prediction researchers and practitioners.  If IFA is strictly restricted to less than 10, CBS+ and EALTR~\footnote{EALTR is about 0.1 less in Recall@20\% score compared to ManualUp according to \cite{yu2024improving}.} are the choices, however, they would lose about 0.1-0.27 in recall of defective modules when compared to ManualUp.  If an IFA less than 15 is acceptable, then EA-Z-UBag-svm can find about 0.8\% defective modules more than ManualUp, and of the order of 7 initial false alarms less than ManualUp.  Developers can choose the predicting methods according to their preferences.  From the perspective that the original goal of effort-aware defect prediction is to improve cost-effectiveness, finding more defects with limited effort is a useful contribution.

\bibliographystyle{ACM-Reference-Format}
\bibliography{sample-base}

\appendix

\end{document}